\documentstyle[aps,prl,cite,enumerate,verbatim]{revtex}
\begin{document}
\title{ Entangled qutrits violate local realism stronger than qubits --an  analytical proof }
\author{Jing-Ling Chen,$^{1,3}$ Dagomir Kaszlikowski,$^{1,2}$ L. C. Kwek,$^{1,4}$ 
Marek \.Zukowski,$^{5}$ and C. H. Oh$^1$}
\address{$^1$Department of Physics, National University of Singapore, 10
Kent Ridge Crescent, Singapore 119260 \\
$^2$Instytut Fizyki Do\'swiadczalnej, Uniwersytet Gda\'nski, PL-80-952, Gda\'nsk, Poland,\\
$^3$ Laboratory of Computational Physics, 
Institute of Applied Physics and Computational Mathematics, \\
P.O. Box 8009(26), Beijing 100088, People's Republic of China \\
$^4$National Institute of Education, Nanyang Technological University, 1 Nanyang Walk, Singapore 639798\\
$^5$Instytut Fizyki
Teoretycznej i Astrofizyki, Uniwersytet Gda\'nski, PL-80-952,
Gda\'nsk, Poland.}

\maketitle

\begin{abstract}

In Kaszlikowski {\it et al.} [Phys. Rev. Lett. {\bf 85}, 4418 (2000)], it has been shown
numerically that the violation of local realism for two maximally
entangled $N$-dimensional ($3 \leq N$) quantum objects is stronger
than for two maximally entangled qubits and grows with $N$. In
this paper we present the analytical proof of this fact for $N=3$.

\end{abstract}


Since the formulation of Bell theorem \cite{BELL}, various forms of so called Bell 
inequalities, see for instance
\cite{INEQUALITIES},
have been devised to investigate the possibility (or lack of such possibility)
of local realistic description of
correlations observed in various quantum systems such as
$M$ entangled $N$ dimensional quantum objects. The main advantage of this approach is
its simplicity. The 
drawback of this method is that, in general, Bell inequalities are only a necessary
condition for the existence of local and realistic description of the investigated
quantum system. Only in a few cases, see for instance \cite{FINE,WERNER,BRUKNER}, 
it has been proved that some Bell inequalities are
necessary and sufficient condition for the existence of local realism. All these cases
deal with two \cite{FINE} or more than two \cite{WERNER,BRUKNER} qubits and two local observables
measured at each side of the Bell experiment.

In \cite{MERMIN} a general approach to the problem has been presented. It 
is possible to find all relevant inequalities that have to be fulfilled by the
probabilities obtained by the measurement of any number of local observables on
the system consisting of an arbitrary number of quantum objects, each of which
described by a Hilbert space of arbitrary dimesnion, so that it can be described in terms
of local realism. However, the number of inequalities that have to be examined grows
extremely fast with the dimension of the problem, i.e., number of local observables, quantum
objects and dimension of Hilbert space describing given objects. This makes the method
practically useless, as shown in \cite{PITOVSKY,PERESALL}.

Recent research shows that a different approach is possible.
In \cite{BATURO, KASZLIKOWSKI} methods of numerical linear optimization has
been successfully applied to two qubit correlations with up to ten local
observables being measured at each side of the experiment \cite{BATURO}
and for two $N$-dimensional objects (2 $\leq N \leq 16$)
with two local observables at each side of the experiment \cite{KASZLIKOWSKI, DURT}.
In this approach one does not find Bell inequalities but finds the conditions under which for
the given quantum system and quantum observables measured on it there exists a local hidden
variable model reproducing quantum results. Additionally, this method can be directly applied to the
analysis of experimental data.

The paper \cite{KASZLIKOWSKI} is a good example of how important it is
to know necessary and sufficient conditions for the existence of local realism in the
given case. For instance in \cite{GISIN}, it was shown for two $N$-dimensional entangled systems
that the Clauser-Horne-Shimony-Holt (CHSH) inequality \cite{CHSH} is maximally
violated by the factor of $\sqrt 2$. The reason for this is that CHSH inequality is not
a sufficient condition for the existence of local realism for two entangled objects
each described by a Hilbert space of the dimension greater than two. Indeed, the results of 
\cite{KASZLIKOWSKI} show that violations of local realism increase with the dimension of the systems.

In this paper we prove analytically that the violation of local
realism for two maximally entangled qutrits (objects described by
a three dimensional Hilbert space) observed via two unbiased three
input and three output beamsplitters 
\cite{TRITTERS} is stronger than for two maximally entangled
qubits \cite{KASZLIKOWSKI}. Earlier numerical computations advocated such a violation but 
rigorous analytical evidence has so far been lacking except for the trivial
case of qubit. Thus it is anticipated that analytical proofs should exist for 
higher dimensional quantum systems. Our present work on qutrit therefore constitutes 
the first such attempt to confirm the previous numerical claim. Moreover, we also
see that the extension from qubit to qutrit is clearly non-trivival. In fact, a 
comparison of our results with the separability
condition for so called generalised Werner states \cite{WERNERSTATES} can shed a new
light on the relation between local realism and separability of bipartite quantum 
systems.

We consider the Bell type experiment in which two spatially separated observers
Alice and Bob measure two
non-commuting observables $A_1,A_2$ for Alice and $B_1,B_2$ for Bob
on the maximally entangled state $|\psi\rangle$ of two qutrits

\begin{eqnarray}
&&|\psi\rangle = {1\over\sqrt 3}(|0\rangle_A |0\rangle_B+
|1\rangle_A |1\rangle_B+|2\rangle_A |2\rangle_B),
\label{state}
\end{eqnarray}
where $|k\rangle_{A}$ and $|k\rangle_{B}$ describe k-th basis state of
the qutrit $A$ and $B$ respectively. Such a state can be prepared
with pairs of photons with the aid of parametric down conversion (see
\cite{TRITTERS}), in which case kets $|k\rangle_{A}$ and $|k\rangle_{B}$
denotes photons propagating to Alice and Bob in mode $k$.   

Here we consider the special case in which both observers measure observables defined
by 6-port (three input and three output ports) beam splitter.
The extended theory of
such devices can be found in \cite{TRITTERS}. Here we give only a brief description.

{\it Unbiased} $6$-port beamsplitter, which is called tritter,
\cite{TRITTERS}
is a device with the following property: if one photon enters into
any single input port (out of the $3$), its chances of exit are
equally split between all $3$ output ports. 
One can always build tritter with the distinguishing trait that
the elements of its unitary transition matrix, $\hat{T}$, are
{\it solely} powers of the $3$-rd root of unity
$\alpha=\exp{(i2\pi/3)},$
namely $T_{kl}= \frac{1}{\sqrt{3}}\alpha^{(k-1)(l-1)}.$
In front of $i$-th input port of the tritter we put a phase shifter that
changes the phase of the incoming photon by $\phi_i$. These three phase
shifts, which we denote for convenience as a ``vector" of phase shifts 
$\vec{\phi}=(\phi_1,\phi_2,\phi_3)$, are macroscopic local parameters
that can be changed by the observer. Therefore, tritter together with the 
three phase shifters performs the unitary transformation 
$\hat{U}(\vec{\phi})$ with the entries $U_{kl}=T_{kl}\exp(i\phi_l)$.

Alice and Bob measure the following observables
\begin{eqnarray}
&&A (\phi_i)= \hat{U}(\vec{\phi}_i)|0\rangle\langle
0|\hat{U}(\vec{\phi}_i)^{\dagger}+ \alpha
\hat{U}(\vec{\phi}_i)|1\rangle\langle
1|\hat{U}^{\dagger}(\vec{\phi_i})+ \alpha
^2\hat{U}(\vec{\phi}_i)|2\rangle\langle
2|\hat{U}^{\dagger}(\vec{\phi}_i)\nonumber\\
&&B(\theta_j)
=\hat{U}(\vec{\theta_j})|0\rangle\langle
0|\hat{U}(\vec{\theta}_j)^{\dagger}+ \alpha
\hat{U}(\vec{\theta}_j)|1\rangle\langle
1|\hat{U}^{\dagger}(\vec{\theta}_j)+ \alpha
^2\hat{U}(\vec{\theta}_j)|2\rangle\langle
2|\hat{U}^{\dagger}(\vec{\theta}_j), 
\label{observables}
\end{eqnarray}
where $i,j=1,2$ and where, for instance, $\vec{\phi}_{i}$ denotes
the vector of local phase shifts for Alice in the $i$-th
experiment. Please notice that we ascribe complex numbers to the
results of measurements, i.e., to the ``click" of the $l$-th detector
we ascribe the number $\alpha^l$. The justification of such an
assigment can be found in \cite{TRITTERS}. It results in a very
symmetrical complex correlation function
\begin{eqnarray}
&&E(\phi_{i},\theta_{j})= \langle\psi|A(\phi_i)
B(\theta_j)|\psi\rangle \nonumber\\ 
&& ={1\over 3}(\exp(\phi^{1}_{i}-\phi^{2}_{i}+\theta^{1}_{j}-\theta^{2}_{j})+
\exp(\phi^{2}_{i}-\phi^{3}_{i}+\theta^{2}_{j}-\theta^{3}_{j})\nonumber\\
&&+\exp(\phi^{3}_{i}-\phi^{1}_{i}+\theta^{3}_{j}-\theta^{1}_{j})),
\end{eqnarray}
where, for instance, $\phi^{1}_{i}$ denotes the first phase shift
at Alice's side in the $i$-th experiment. This correlation
function retains the information about the correlations observed
in the experiment. In fact, according to quantum mechanics the
whole information that is accessible in the experiment are
probabilities of coincidence firings of the detectors. It can be
easily verified through the knowledge of the correlation function
$E(\phi_{i},\theta_{j})$ one is able to calculate the
probabilities of these coincidence ``clicks" and in this way obtain
the whole information about the correlations observed in the
system.

Following \cite{KASZLIKOWSKI} We define the strength of violation of local
realism as the minimal noise admixture $F_{thr}$ to the state (\ref{state}) 
below which the measured correlations cannot be described by local
realism for the given observables. Therefore, we assume that Alice
and Bob perform their measurements on the following mixed state
$\rho_{F}$
\begin{eqnarray}
&&\rho_{F}=(1-F)|\psi\rangle\langle\psi|+F\rho_{noise},
\label{noise}
\end{eqnarray}
where $0 \leq F \leq 1$ and where $\rho_{noise}$ is a diagonal
matrix with entries equal to $1/9$. This matrix is a totally
chaotic mixture (noise), which admits a local and
realistic description. For $F=0$ (pure maximally entangled state)
local realistic description does not exist whereas for $F=1$ (pure
noise) it does. Therefore, there exists some threshold value of
$F$, which we denote by $F_{thr}$, such that for every $F\leq
F_{thr}$ local and realistic description does not exist. The bigger the value
of $F_{thr}$, the stronger is the violation of local realism. The
correlation function for the state (\ref{noise}) reads
$E^{F}(\vec{\phi}_i,\vec{\theta}_j)=(1-F)E(\vec{\phi}_i,\vec{\theta}_j)$.

Let us now assume that Alice measures two observables defined by
the following sets of phase shifts
$\vec{\phi}_1=(0,\pi/3,-\pi/3),\vec{\phi}_2=(0, 0, 0)$ whereas Bob measures two
observables defined by the sets of phase shifts
$\vec{\theta}_1=(0, \pi/6, -\pi/6),\vec{\theta}_2=(0, -\pi/6, \pi/6)$. From numerical
computations it is known \cite{KASZLIKOWSKI, DURT} that these sets
of phases gives the highest $F_{thr}$. Straightforward calculations
give the following values of the correlations functions for each
experiment:
$E^F(\vec{\phi}_1,\vec{\theta}_1)=E^F(\vec{\phi}_2,\vec{\theta}_2)=Q_1=\frac{2\sqrt
3 +1}{6}-i\frac{2-\sqrt
3}{6},E^F(\vec{\phi}_1,\vec{\theta}_2)=Q_{1}^{*},E^F(\vec{\phi}_2,\vec{\theta}_1)=Q_2=
-{1\over 3}(1+2i)$. From these complex numbers we can create a
$2\times 2$ matrix $\hat{Q}$ with the entries
$(\hat{Q})_{ij}=E^F(\vec{\phi}_i,\vec{\theta}_j)$.

Local realism implies the following 
structure of the correlation function that is to reproduce 
the quantum correlation function defined above
\begin{equation} 
E_{LHV}(\vec{\phi}_i, \vec{\theta}_j)= 
\int d\lambda\rho(\lambda)A(\vec{\phi}_i,\lambda)B(\vec{\theta}_j,\lambda), 
\end{equation}
where for trichotomic measurements $A(\vec{\phi}_i,\lambda)=\alpha^m$ and $B(\vec{\theta}_j,\lambda)=\alpha^n$,
$(m, n=1, 2, 3)$. Three-valued functions $A(\vec{\phi}_i,\lambda),B(\vec{\theta}_j,\lambda)$
represent the values of local
measurements predetermined by local hidden variables, denoted by $\lambda$, for the
specified local settings. This expression is an average over a certain
local hidden variable distribution $\rho(\lambda)$ of certain {\it factorisable} matrices, 
namely those with elements given by
$H^{ij}_{\lambda}=A(\vec{\phi}_i,\lambda)B(\vec{\theta}_j,\lambda)$.  The
symbol $\lambda$ may hide very many parameters.  However, since the
only possible values of $A(\vec{\phi}_i,\lambda)$ and
$B(\vec{\theta}_j,\lambda)$ are $1,\alpha,\alpha^2$ there are only $9$ {\it
different} sequences of the values of $(A(\vec{\phi}_1,\lambda),
A(\vec{\phi}_2,\lambda))$, and $9$ {\it different} sequences of the values of
$(B(\vec{\theta}_1,\lambda),B(\vec{\theta}_2,\lambda))$,
and consequently they form only $81$ matrices $\hat{H}_{\lambda}$. 

Therefore the structure of local hidden variable model of $E_{LHV}(\vec{\phi}_i,
\vec{\theta}_j)$ reduces to discrete probabilistic model involving the
average of all the $81$ matrices $\hat{H}_{\lambda}$. Therefore, we replace
the parameter $\lambda$ by index $k$ ($k=1,2,\dots,81$)
to which we ascribe the matrix $\hat{H}_k$ with entries 
$H^{ij}_k=\alpha^{k_i+l_j}$ ($i,j=1,2$), where 
$k_1=[(k-1)/9]-1,k_2=[(k-1)/3]-1,l_1=1,l_2=k$ (please notice
that $\alpha^{-1}=\alpha^2$) and where $[x]$ denotes the integer part of the number $x$. 
It can be checked that only first $27$ matrices are different, which means
that it suffices to consider only them. With this notation the correlation function
$E_{LHV}(\vec{\phi}_i, \vec{\theta}_j)$ acquires the following simple form
\begin{equation} 
  E_{LHV}(\vec{\phi}_i, \vec{\theta}_j)=
  \sum_{k=1}^{27}p_{k}H^{ij}_k,
  \label{MODEL}
\end{equation}
with, of course, the probabilities satisfying $p_{k}\geq0$ and
$\sum_{k=1}^{27}p_{k}=1$. From $E_{LHV}(\vec{\phi}_i, \vec{\theta}_j)$
we build the matrix $\hat{E}_{LHV}$.

Quantum predictions in form of the matrix $\hat{Q}^{F}$ can be recovered
by local hidden variables if and only if

\begin{eqnarray}
&&\hat{Q}^{F}=\sum_{n=1}^{27} p_{n}\hat{H}_{n}.
\label{matrixform}
\end{eqnarray}
Now, we want to find the minimal possible $F$ for which it is still
possible to recover matrix $\hat{Q}^{F}$ using the
probability distribution $p_{n}$ and factorizable matrices
$\hat{H}_{n}$. For convenience we define a new parameter $V=1-F$. Then
the minimal $F$ refers to maximal $V$.

Theorem: The maximal $V$ equals $V_{thr}={6\sqrt 3 - 9 \over 2}$.

Proof: First we observe that matrix $\hat{Q}^{V}$ can be written in the following way
\begin{eqnarray}
\label{e2} \hat{Q}^{V}= V \biggr[ \frac{2\sqrt 3
+1}{6}-i\frac{2-\sqrt 3}{6} \biggr]I + V \biggr[ -
\frac{2\sqrt{3}-1}{6} + i \frac{2+\sqrt{3}}{6} \biggr]
{\vec{n}}\cdot {\vec \sigma},
\end{eqnarray}
where ${\vec n}=(n_x, n_y, n_z)=\biggr( -\frac{1}{2},
\frac{\sqrt{3}}{2},0 \biggr)$ (please notice that $|{\vec n}|=1$).
One observes that $\hat{Q}^{V}$ commutes with the matrix
$\hat{\cal U}={\vec{n}}\cdot {\vec \sigma}$, which has only two nonzero
entries ${\cal U}_{12}=\alpha^2,{\cal U}_{21}=\alpha$ and is unitary and
hermitian. Furthermore, $\hat{\cal U}$ preserves the structure of
matrices $\hat{H}_{n}$ in the sense that for every $n=1,2,\dots,27$,
$\hat{\cal U}\hat{H}_{n}\hat{\cal U}=\hat{H}_{m}$ for some $m=1,2,\dots,27$. This is one
to one mapping. One can also find that some matrices $\hat{H}_{n}$ are
invariants with respect to transformation $\hat{\cal U}$. For further
considerations it is necessary to have the list of pairs $(n,m)$
\begin{eqnarray}
\{(1, 8),(2, 10),(3, 24),(4, 17),(5, 19),(7, 26),(9, 15),(11, 13),
(12, 27),(14, 22),(16, 20),(23, 25)\}. \label{pairs}
\end{eqnarray}
By considering the $i$-th pair $(n,m)$ in the above list, we can define new 
matrices, $\hat{G}_i = \hat{H}_n + \hat{H}_m$.  Thus, $\hat{G}_1=\hat{H}_1 + \hat{H}_8$, 
$\hat{G}_2 = \hat{H}_2 + \hat{H}_{10}$ 
and so forth.
The remaining invariant matrices are $\hat{H}_6, \hat{H}_{18}, \hat{H}_{21}$.
For convenience, we label them as $\hat{G}_{13}$ to $\hat{G}_{15}$, viz 
$\hat{G}_{13}=\hat{H}_6, \hat{G}_{14}=\hat{H}_{18}, \hat{G}_{15}=\hat{H}_{21}$.

Suppose that we have the optimal solution (the solution for which
$V=V_{thr}$ ), i.e., we have the probability distribution $p_{n}$
so that $\hat{Q}^{V_{thr}}=\sum_{n=1}^{27}p_{n}\hat{H}_{n}$. Acting on
both sides of this equation with matrix $\hat{\cal U}$ we get another
optimal solution with the same $V_{thr}$ (matrix $\hat{\cal U}$ commutes
with $\hat{Q}^{V_{thr}}$) but with the new probability distribution
$p'_{k}$, which can be obtained from the previous one by
swapping probabilities belonging to the same pair, for instance,
$p'_{10}=p_2$ and so on. Therefore, due to the above property,
we can assume {\it without loosing generality} that in the optimal
solution the probabilities referring to the same pair are equal.
Therefore, we have reduced the number of relevant probabilities
from $27$ to $15$. One can observe that
{\it every} matrix $\hat{G}_{k}$ can be expressed by matrices $\hat{G}_1,
\hat{G}_{10}, \hat{G}_{13}$ by multiplying them by $\alpha$, $\alpha^2$ and
$-1$. For instance, $\hat{G}_6=\alpha \hat{G}_{10}, \hat{G}_{11}=-\hat{G}_{13}$ etc. Three
matrices are the same: $\hat{G}_5=\hat{G}_3,\hat{G}_7=\hat{G}_4,\hat{G}_{11}=\hat{G}_9$, which further
reduces the number of relevant probabilities from $15$ to $12$.

Having in mind the above properties we can write the optimal
solution in the new form
\begin{eqnarray}
\hat{Q}^{V_{thr}}=\sum_{k\neq 5, 7, 11}w_{k}\hat{G}_{k},
\end{eqnarray}
remembering that now the normalization condition for probabilities $w_k$ reads
\begin{eqnarray}
 2( w_1+ w_2+ w_3+w_4 + w_6+ w_8+w_9+w_{10}+w_{12})+ w_{13}+w_{14}+w_{15}=1.
\end{eqnarray}
Due to the fact that all $\hat{G}_k$ can be expressed by $\hat{G}_1$,
$\hat{G}_{10}$, $\hat{G}_{13}$, we have
\begin{eqnarray}
&&\hat{Q}^{V_{thr}}=( w_1 + \alpha w_8 + \alpha^2 w_{12}) \hat{G}_1 +
(w_{10} + \alpha w_6 + \alpha^2 w_2) \hat{G}_{10}\nonumber\\
&&+[(w_{13}-w_9) + \alpha (w_{14}-w_3) + \alpha^2 (w_{15}-w_4)]
\hat{G}_{13}.
\end{eqnarray}
Notice that $\hat{G}_1+ \hat{G}_{10} -\hat{G}_{13}=0$ so that we have
\begin{eqnarray}
&&\hat{Q}^{V_{thr}}=[(w_1+ w_{13}-w_9) + \alpha(w_8 +w_{14}-w_3)+
 \alpha^2 (w_{12}+ w_{15}-w_4)] \hat{G}_1 \nonumber\\
&& +[(w_{10}+ w_{13}-w_9) + \alpha(w_6 +w_{14}-w_3)+
 \alpha^2 (w_2+ w_{15}-w_4)] \hat{G}_{10}.
\label{3}
\end{eqnarray}
Matrix $\hat{G}_1$ (with entries $G_{1}^{11}=2$, $G_{1}^{12}=-\alpha^2$, $G_{1}^{21}=-\alpha$,
$G_{1}^{22}=2$) is a sum of matrices $\hat{H}_1,\hat{H}_8$ whereas matrix
$\hat{G}_{10}$ (with entries $G_{10}^{11}=-1$, $G_{10}^{12}=2\alpha^2$, $G_{10}^{21}=2\alpha$,
$G_{10}^{22}=-1$) is a sum of matrices $\hat{H}_{14},\hat{H}_{22}$. These four matrices
are linearly independent so they form a basis in four dimensional
space of $2\times 2$ complex matrices. The expansion of
$\hat{Q}^{V_{thr}}$ in this basis reads
\begin{eqnarray}
&&\hat{Q}^{V_{thr}}=\lambda_1\hat{G}_1+\lambda_{10}\hat{G}_{10}, \label{5}
\end{eqnarray}
where $\lambda_1= V_{thr} [ ( \frac{1}{6}+
\frac{1}{3\sqrt{3}})+i ( -\frac{1}{9} +
\frac{1}{2\sqrt{3}})],
 \lambda_{10}= V_{thr} [ ( \frac{1}{6}- \frac{1}{3\sqrt{3}})
 + i ( \frac{1}{9} + \frac{1}{2\sqrt{3}})]$.
Because both $\lambda_1$ and $\lambda_{10}$ lie on the complex
plane between complex numbers $1$ and $\alpha$, they can be
uniquely expressed by these numbers 1 and $\alpha$ with positive
coefficients, i.e., $\lambda_1= V_{thr}[\frac{1}{27}( 9 +
2\sqrt{3}) + \alpha \; \frac{1}{27}( 9 - 2\sqrt{3})]$ and
$\lambda_{10}=V_{thr} [ \frac{1}{27}( 9 - 2\sqrt{3}) + \alpha
\; \frac{1}{27}( 9 + 2\sqrt{3})]$.

We can rewrite the formula (\ref{3}) using the indentity
$1+\alpha+\alpha^2=0$ in the following form
\begin{eqnarray}
 \lambda_1= (w_1+ w_4 + w_{13}-w_9-w_{12}-w_{15}) +
               \alpha(w_4+w_8 +w_{14}-w_3-w_{12}-w_{15}),\nonumber\\
 \lambda_{10}= (w_4+w_{10}+ w_{13}-w_2-w_9-w_{15})
 + \alpha (w_4+w_6 +w_{14}-w_2-w_3-w_{15}).
\label{4}
\end{eqnarray}
After comparing Eq.(\ref{4}) and Eq.(\ref{5}) we have
\begin{eqnarray}
&&w_1+ w_4 + w_{13}-w_9-w_{12}-w_{15}= \frac{V_{thr}}{27}( 9 + 2\sqrt{3}),\nonumber\\
&&w_4+w_8 +w_{14}-w_3-w_{12}-w_{15}= \frac{V_{thr}}{27}( 9 - 2\sqrt{3}),\nonumber\\
&&w_4+w_{10}+ w_{13}-w_2-w_9-w_{15} =\frac{V_{thr}}{27}( 9 - 2\sqrt{3}),\nonumber\\
&&w_4+w_6 +w_{14}-w_2-w_3-w_{15}=\frac{V_{thr}}{27}( 9 + 2\sqrt{3})
\label{e12}
\end{eqnarray}

Because we deal with the optimal solution for which $V$ is maximal
$V=V_{thr}$ all the probabilities with negative sign in (\ref{e12})
must be zero (please notice that none of the probabilities that
come into (\ref{e12}) with negative sign appears in any equation
with a positive sign). We get
\begin{eqnarray}
&&w_1+ w_4 + w_{13}= \frac{V_{thr}}{27}( 9 + 2\sqrt{3}),\nonumber\\
&& w_4+w_8 +w_{14}= \frac{V_{thr}}{27}( 9 - 2\sqrt{3}),\nonumber\\
&& w_4+w_{10}+ w_{13}= \frac{V_{thr}}{27}( 9 -
2\sqrt{3}),\nonumber\\ &&w_4+w_6 +w_{14}=\frac{V_{thr}}{27}( 9 +
2\sqrt{3}).
\end{eqnarray}
Now the whole probability distribution consists of $w_1, w_4, w_6, w_8, w_{10}, w_{13}, w_{14}$.
By substracting the fourth equation from the second one and the third one from the first one we arrive at
\begin{eqnarray}
w_6-w_8={4\sqrt3\over 27}V_{thr}, \;\;\; w_1-w_{10}= {4\sqrt3\over
27}V_{thr}.
\end{eqnarray}
Again, because $V_{thr}$ is maximal, it must be $w_8=w_{10}=0$.
Thus $w_1=w_6={4\sqrt3\over 27}V_{thr}$ and the second and the
third equation in (\ref{e12}) become
\begin{eqnarray}
\label{e16}
w_4+w_{14}= \frac{V_{thr}}{27}( 9 - 2\sqrt{3}), \;\;\;
w_4+w_{13}= \frac{V_{thr}}{27}( 9 - 2\sqrt{3}),\nonumber\\
\end{eqnarray}
This clearly implies $w_{13}=w_{14}=q$. Normalization condition
now reads
\begin{eqnarray}
2( w_1+ w_4 + w_6)+ w_{13}+w_{14}=1.
\end{eqnarray}
A simple algebra gives
\begin{eqnarray}
&&q+w_4={1\over 2}-{8\sqrt3\over 27}V_{thr}.
\end{eqnarray}
However, from (\ref{e16}), we know that $q+w_4={V_{thr}\over
27}(9-2\sqrt3)$. Therefore

\begin{eqnarray}
{1\over
2}-{8\sqrt3\over 27}V_{thr}={V_{thr}\over 27}(9-2\sqrt3)
\end{eqnarray}
which gives $V_{thr}={6\sqrt3 - 9\over 2}$. This ends the proof.

We have shown analytically that for the Bell experiment with the
four trichotomic observables (\ref{observables}) (two at each side
of the experiment) defined by the sets of phase shifts
$\vec{\phi}_1=(0,\pi/3,-\pi/3)$,
$\vec{\phi}_2=(0,0,0)$,
$\vec{\theta}_1=(0, \pi/6, -\pi/6)$,
$\vec{\theta}_2=(0, -\pi/6, \pi/6)$ 
the minimal noise
admixture $F_{thr}$ above which local and realistic description
exists is $F_{thr}=1-V_{thr}={11-6\sqrt 3\over 2}$. For two maximally
entangled qubits this number is ${2-\sqrt 2\over 2} < F_{thr}$.
Therefore, two entangled qutrits are more robust against local and
realistic description than two entangled qubits.

Although, the presented here proof cannot be easily applied to the
set of arbitrary observables defined in (\ref{observables}) as it
relies on the symmetry properties of matrix $\hat{Q}^{F}$ it may
be considered as the first step towards the Bell theorem for two
entangled qutrits.

MZ thanks Anton Zeilinger and Alipasha Vaziri for discussions.
MZ and DK are supported by the University of Gdansk Grant No.
BW/5400-5-0032-0 and by KBN grant No. 5 P03B 088 20. This 
paper is also supported in part under NUS research grant No. R-144-000-054-112.


\begin{references}

\bibitem{BELL} J. Bell, Physics {\bf 1}, 195 (1964).

\bibitem{INEQUALITIES} J. F. Clauser, M. A. Horne, A. Shimony and R. A. Holt, 
Phys. Rev. Lett. {\bf 23}, 15, 880 (1969); 
E. P. Wigner, Am. J. Phys. {\bf 38}, 8, 1005 (1970);
J. F. Clauser and M. A. Horne, Phys. Rev. D {\bf 10}, 526 (1974);
N. D. Mermin, Phys. Rev. D {\bf 22}, 2, 356 (1980); 
S. L. Braunstein and C. M. Caves, Ann. Phys. (NY) {\bf 202}, 22 (1990).
N. D. Mermin, Phys. Rev. Lett. {\bf 65}, 1838 (1990);
M. Ardehali, Phys. Rev. D. {\bf 44}, 10, 3336 (1991); 
N. Gisin and A. Peres, Phys. Lett. A. {\bf 162},1,15 (1992);
A. V. Belinskii, D. N. Klyshko, Phys. Usp. {\bf 36}, 653 (1993);
M. \.Zukowski, Phys. Lett. A {\bf 177},4-5,290 (1993);
N. Gisin, H. Bechmann-Pasquinucci, Phys. Lett. A {\bf 246}, 1 (1998);
N. Gisin, Phys. Lett. A {\bf 260}, 1 (1999);
D. Kaszlikowski and M. \.Zukowski, Phys. Rev. A {\bf 61},2, 022114 (2000);
I. Pitovsky, K. Svozil, quant-ph//0011060.

\bibitem{FINE} A. Fine, Phys. Rev. Lett. {\bf 48}, 291 (1982).

\bibitem{WERNER}  R. F. Werner, M. M. Wolf, quant-ph//0102024.

\bibitem{BRUKNER} M. \.Zukowski and C. Brukner, quant-ph//0102039.

\bibitem{MERMIN} N. D. Mermin and G. Schwarz, Found. Phys. {\bf 12},101 (1982).

\bibitem{PITOVSKY} I. Pitovsky, Math. Programming {\bf 50}, 395 (1991).

\bibitem{PERESALL} A. Peres, Found. Phys. {\bf 29}, 589 (1999).

\bibitem{BATURO} M. \.Zukowski, D. Kaszlikowski, A. Baturo, J-A.
Larsson, quant-ph//9910058.

\bibitem{KASZLIKOWSKI} D. Kaszlikowski, P. Gnaci\'nski, M. \.Zukowski, W. Miklaszewski and
A. Zeilinger, Phys. Rev. Lett. {\bf 85}, 4418 (2000).

\bibitem{DURT} T. Durt, D. Kaszlikowski and M. \.Zukowski,
quant-ph//0101084.

\bibitem{GISIN} N. Gisin and A. Peres, Phys. Lett. A {\bf 162}, 15 (1992).

\bibitem{CHSH} J. F. Clauser, M. A. Horne, A. Shimony and R. A. Holt,
Phys. Rev. Lett. {\bf 23}, 880 (1969).

\bibitem{TRITTERS} M. \.Zukowski, A. Zeilinger, M. A. Horne, Phys. Rev. A {\bf 55}, 
2564 (1997).

\bibitem{WERNERSTATES} M. Horodecki and P. Horodecki, Phys. Rev. A {\bf 59}, 4206 (1999).
\end{references}
\end{document}